\documentclass[conference]{IEEEtran}
\usepackage[utf8]{inputenc}
\usepackage[T1]{fontenc} 
\usepackage{gensymb} 
\usepackage{array}
\usepackage{amsmath}
\usepackage{graphicx}
\usepackage{cite}
\usepackage[colorlinks=true, linkcolor=blue, citecolor=blue, urlcolor=blue]{hyperref}
\usepackage[numbers]{natbib} 
\usepackage[left=.75in, right=.75in, top=1in, bottom=.75in]{geometry} 
\makeatletter

\@ifundefined{showcaptionsetup}{}{
 \PassOptionsToPackage{caption=false}{subfig}}
\usepackage{subfig}
\makeatother

\usepackage{eso-pic}

\begin{document}

\title{Design and Optimization of a Metamaterial Absorber for Solar Energy Harvesting in the THz Frequency Range}

\author{\IEEEauthorblockN{Nafisa Anjum \IEEEauthorrefmark{1}, and Alok Kumar Paul \IEEEauthorrefmark{2}}
\IEEEauthorblockA{\IEEEauthorrefmark{1}\IEEEauthorrefmark{2}Dept. of Electrical \& Electronic Engineering (EEE), \\Rajshahi University of Engineering \& Technology, Bangladesh}
\IEEEauthorblockA{Emails: nafisaanjum9999@gmail.com,  alok@eee.ruet.ac.bd}
}

\maketitle

\begin{abstract}

This research highlights the design and comprehensive analysis of a high-performance three-layer metamaterial absorber tailored for solar energy harvesting in the terahertz (THz) frequency range.  The proposed absorber incorporates gold (Au), vanadium dioxide (VO\textsubscript{2}), and silicon dioxide (SiO\textsubscript{2}) to attain near-unity absorption throughout an extensive spectral range.  The design is optimized to utilize the distinctive optical and plasmonic characteristics of these materials, facilitating effective solar radiation absorption in the THz range.  The absorber has a wide absorption bandwidth of 3.00 THz, spanning frequencies from 2.414 THz to 5.417 THz, with peak absorption approaching 99\%.  This range renders it an exceptionally effective alternative for improving solar energy conversion.  The absorber exhibits polarization insensitivity, guaranteeing consistent performance at different incidence angles.  Theoretical simulations were conducted to optimize the metamaterial's geometric configuration and material composition, highlighting the essential function of SiO\textsubscript{2} as a dielectric spacer and the adjustable characteristics of VO\textsubscript{2}, which facilitate dynamic absorption regulation.  These findings underscore the absorber's potential for advanced solar energy harvesting devices.

\end{abstract}


\section{Introduction}

Artificial engineered composites known as metamaterials exhibit distinctive electromagnetic properties absent in natural materials. These materials offer surpassed control over electromagnetic waves, acoustic signals, and several physical events; their structural arrangement rather than their fundamental composition defines them most. Metamaterials have drawn a lot of attention in several fields, including optics, telecommunications, and acoustics since its theoretical conception and later real application in the late 20th and early 21st centuries. \cite{pendry2000negative}.

Veselago proposed materials with negative refractive indices, implying that they might reverse Snell's Law, hence laying the theoretical foundation for metamaterials. Later experimental confirmation of this theory came from Smith \textit{et al.}, who effectively showed negative refraction in synthetic structured metamaterials \cite{smith2000composite}. Their groundbreaking work marked a turning point in the discipline by offering concrete proof that built objects may control electromagnetic radiation at microwave frequencies \cite{smith2000composite}.

Development of extremely effective electromagnetic absorbers is one of the most important applications of metamaterials. These absorbers are made to gather and dissipate incident electromagnetic waves within specified frequency bands. They accomplish this by optimising impedance matching via sub-wavelength resonant structures, hence reducing reflection and maximising absorption \cite{landy2008perfect}. The exact geometric arrangement of the elements in a metamaterial absorber determines its performance mostly in providing almost 100\% absorption and wide operational bandwidth.

Metamaterial-based solar absorbers have emerged as a promising technology for enhancing solar energy harvesting. By leveraging carefully engineered nanostructures, these absorbers demonstrate superior efficiency in trapping and converting solar radiation across a wide spectral range \cite{padilla2006dynamical}. Among the several operational frequency ranges, the terahertz (THz) spectrum—which spans approximately 0.1 to 10 THz—has attracted much attention. Operating in this spectrum, metamaterials show remarkable wave-matter interactions that allow very effective absorption. THz metamaterial absorbers are thus quite useful for sensing, imaging, and energy harvesting \cite{padilla2006dynamical}.

This work presents a broadband metamaterial absorber intended especially for terahertz solar energy harvesting. The proposed absorber has a multi-layered design consisting of vanadium dioxide (VO\textsubscript{2}), gold (Au), and silicon dioxide (SiO\textsubscript{2}). These components are laid deliberately to maximise absorption capacity. This work attempts to confirm the feasibility and practical implementation of high-performance metamaterial absorbers in THz applications by means of thorough numerical simulations and theoretical studies.

\section{Literature Review}

Throughout the last twenty years, various ideas for THz metamaterial absorbers have been presented.  In 2002, Gay-Balmaz and Martin developed a single split-ring resonator (SSR) operating at around 1 GHz, illustrating the ability of metamaterials to control electromagnetic waves at microwave frequencies. \cite{gay2002electromagnetic}. Landy \textit{et al.} (2008) proposed a metamaterial absorber, achieving over 88\% absorption, by combining two metallic resonators with a substrate \cite{landy2008perfect}.  This design illustrated the attainment of near-perfect absorption through the combination of electric and magnetic resonances.

In 2011, Shen \textit{et al.} introduced a complementary circular resonator (CCR) exhibiting absorptions of 99\%, 93\%, and 95\% at frequencies of 4.06 GHz and 6.73 GHz.  They also highlighted that their design was polarization-insensitive, enhancing its versatility for practical applications. \cite{shen2011polarization}. In 2013, Pang \textit{et al.} introduced a dual-band metamaterial absorber operating at GHz, utilizing wire architectures. \cite{pang2013analysis}. Two years later, in 2015, Lim \textit{et al.} developed a dual-band absorber utilizing a grounded magnetic substrate with a split-ring resonator (SRR), enhancing absorption efficiency in the GHz range \cite{lim2015dual}. Significant advancements have been accomplished in the terahertz domain. In 2018, Song \textit{et al.} introduced an innovative method using vanadium dioxide (VO\textsubscript{2})-based metamaterials to achieve high-efficiency terahertz absorption. By modifying the conductivity of VO\textsubscript{2}, they developed resonant absorbers capable of functioning across a broad spectrum of terahertz frequencies. Their design attained near-perfect absorption under normal incidence \cite{song2018broadband}. 

In 2021, Wu \textit{et al.} presented a more advanced VO\textsubscript{2}-based metamaterial perfect absorber (MPA) with ultra-wideband terahertz absorption. The ability to alter VO\textsubscript{2} conductivity allowed for variable efficiency. Their design enabled peak intensity modulation ranging from 4\% to 100\% \cite{wu2021ultra}. 

Recent studies have explored alternative materials and design approaches to enhance absorption efficiency and tunability. Elkorany \textit{et al.} \cite{elkorany2023design} introduced a metallic metamaterial absorber that achieved perfect absorption. However, its narrow bandwidth limited its use in broadband THz applications. Similarly, Asgari \textit{et al.} \cite{asgari2024multi} proposed a multi-band graphene-based anisotropic absorber. While it achieved near-perfect absorption in certain modes, it showed inconsistencies across different frequency ranges. 

Graphene-based designs have also been investigated for tunability. Ri \textit{et al.} \cite{ri2024tunable} demonstrated a complementary split-ring resonator-based graphene absorber. Their design provided tunability but suffered from a relatively limited bandwidth. Zhang \textit{et al.} \cite{zhang2024graphene} introduced a dual-band graphene-metal-dielectric absorber with frequency and amplitude tunability, though it lacked the continuous broadband absorption required for certain applications. Xie \textit{et al.} \cite{xie2021tunable} leveraged fractal metasurfaces to design a broadband THz absorber, but the reported bandwidth was still limited. Chen \textit{et al.} \cite{chen2020broadband} presented a tunable THz metamaterial absorber using a complementary gammadion-shaped graphene structure, which provided a reasonable trade-off between tunability and bandwidth but required geometric modifications for tuning.

Despite these advancements, several challenges remain. Many designs are hindered by their intricate configurations, large sizes, or lower levels of absorption in specific frequency bands. Limited bandwidth continues to be a significant issue, along with the reliance on expensive materials and complex fabrication processes. Addressing these challenges is crucial for the further development of practical, cost-effective metamaterial absorbers.

\begin{table}[h]
    \caption{Parameter List of the Model}
    \centering
    \begin{tabular}{|c|c|p{5cm}|}
        \hline
        \textbf{Name} & \textbf{Value (\textmu m)} & \textbf{Description} \\
        \hline
        P & 30 & Period of the metasurface unit cell \\
        \hline
        h & 10 & Thickness of the SiO\textsubscript{2} dielectric layer \\
        \hline
        L\textsubscript{1} & 18 & Outer length of the square top layer \\
        \hline
        L\textsubscript{2} & 9 & Inner length of the square top layer \\
        \hline
        t\_Au & 2 & Thickness of the gold ground plane \\
        \hline
        t\_VO\textsubscript{2} & 0.2 & Thickness of the VO\textsubscript{2} top layer \\
        \hline
        Gm & $5.75 \times 10^{13}$ & Collision frequency of VO\textsubscript{2} in Drude model \\
        \hline
        Sg & $2 \times 10^{5}$ & Electrical conductivity of VO\textsubscript{2} \\
        \hline
        Sg0 & $3 \times 10^{5}$ & Parameter related to VO\textsubscript{2} Drude model \\
        \hline
        wp0 & $1.4 \times 10^{15}$ & Plasma frequency parameter in the Drude model \\
        \hline
    \end{tabular}
    \label{table:metasurface_params}
\end{table}

\begin{figure}[htbp]
\centerline{\includegraphics[width=\linewidth]{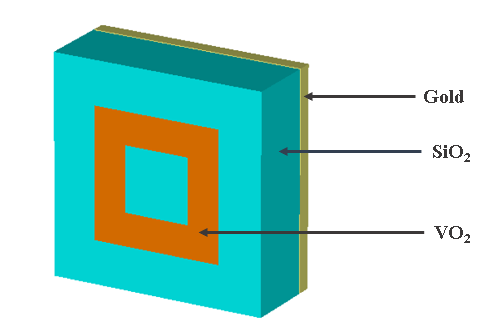}}
\caption{3D schematic view of the proposed metamaterial absorber model.}
\label{fig}
\end{figure}

\section{Methodology}

This research presents a metamaterial absorber with a three-layered architecture, comprising a gold (Au) ground plane, a silicon dioxide (SiO\textsubscript{2}) dielectric spacer, and a vanadium dioxide (VO\textsubscript{2}) upper layer.  Each component has been selected carefully for its electromagnetic properties, enhancing the absorber's performance. VO\textsubscript{2} is crucial because of its capacity to undergo phase transitions, enabling it to alternate between insulating and metallic states in reaction to temperature changes.  This characteristic renders VO\textsubscript{2} extremely adjustable and hence ideal for adaptive metamaterial applications requiring regulated electromagnetic responses.  Its nonlinear characteristics further augment its efficacy in the terahertz (THz) frequency region, where accurate manipulation of electromagnetic waves is essential.  SiO\textsubscript{2} functions as the dielectric spacer, exhibiting low-loss dielectric properties that reduce energy dissipation while delivering thermal and chemical stability, thereby ensuring the absorber's  durability.  The Au ground plane operates as an ideal conductor, inhibiting electromagnetic wave transmission while enhancing energy absorption and reflection.

The absorber features a periodic unit cell with a period of $P = 30~\mu\text{m}$. The dimensions of the VO\textsubscript{2} top layer are $L_1 = 18~\mu\text{m}$ and $L_2 = 9~\mu\text{m}$, with a thickness of $t_{\text{VO}_2} = 0.2~\mu\text{m}$. The Au ground plane has a thickness of $t_{\text{Au}} = 2~\mu\text{m}$ and an electrical conductivity of $4.56 \times 10^{7}$ S/m. The dielectric spacer, composed of SiO\textsubscript{2}, has a thickness of $h = 10~\mu\text{m}$ and a relative permittivity of $\varepsilon = 3.8$.

\subsection{Electromagnetic Wave Propagation and Absorption Mechanism}

The functionality of the metamaterial absorber is determined by the interaction between incident electromagnetic waves and the multilayered structure.The total energy of an incident wave is partitioned into three components: reflection ($R$), transmission ($T$), and absorption ($A$). The transmission component is negligible ($|S_{21}| = 0$) due to the presence of the Au ground plane, which completely blocks wave propagation through the structure. The absorption can be expressed as:

\begin{equation}
|A(\omega)| = 1 - |S_{11}|^2
\end{equation}

where $S_{11}$ denotes the reflection coefficient. Optimal absorption is achieved when reflection is minimized, ensuring near-total energy absorption through impedance matching with free space.

\subsection{Plasma Frequency and Drude Model}

Plasma frequency ($\omega_p$) is a fundamental parameter that characterizes the electromagnetic response of metamaterials.  It denotes the rate at which free electrons oscillate in reaction to an external electromagnetic field. The plasma frequency for metamaterials can be calculated using the equation:

\begin{equation}
\omega_p = \sqrt{\frac{Ne^2}{m\varepsilon_0}}
\end{equation}

where $e$ is the electron charge, $m$ is the electron mass, $N$ is the carrier concentration, and $\varepsilon_0$ is the permittivity of free space.

\subsection{Impedance Matching Theory}

Impedance matching is crucial for maximizing absorption and reducing reflection, ensuring the metamaterial's impedance matches with that of free space.  Specifically, the permittivity ($\varepsilon$) and permeability ($\mu$) of the metamaterial are the factors that define the intrinsic impedance ($\eta$) of the metamaterial. Additionally, the following equation can be used to determine the intrinsic impedance:

\begin{equation}
\eta = \sqrt{\frac{\mu}{\varepsilon}}
\end{equation}

By adjusting material properties and geometrical configurations, near-perfect impedance matching can be achieved, enhancing absorption efficiency across a broad frequency range in the THz spectrum.

\subsection{Design Considerations for Tunability}

The phase transition characteristic of VO\textsubscript{2} facilitates dynamic tunability of the absorber's electromagnetic response.  By adjusting its conductivity through temperature fluctuations, the reflection and absorption properties can be modified instantaneously.  This adjustable behavior is beneficial for applications necessitating adaptive control, including reconfigurable sensing, imaging, and stealth technologies.  The combination of tunable VO\textsubscript{2}, stable SiO\textsubscript{2}, and highly conductive Au produces an absorber that sustains high efficiency throughout an extensive frequency range, especially in THz applications.

\section{Results and Discussion}
This section highlights the simulation outcomes and examines the influence of various structural parameters on the absorber's efficiency.  The Finite Element Method (FEM) is the simulation approach employed to assess the suggested model.  This method facilitates an in-depth examination of each model's behaviour to parameter fluctuations, yielding insights into structure dynamics and possible enhancements or optimizations.  This rigorous computational approach seeks to clarify the practical and theoretical implications of structural modifications within the reviewed models.

\subsection{Simulation Results}

\begin{figure}[htbp]
\centerline{\includegraphics[width=.9\linewidth]{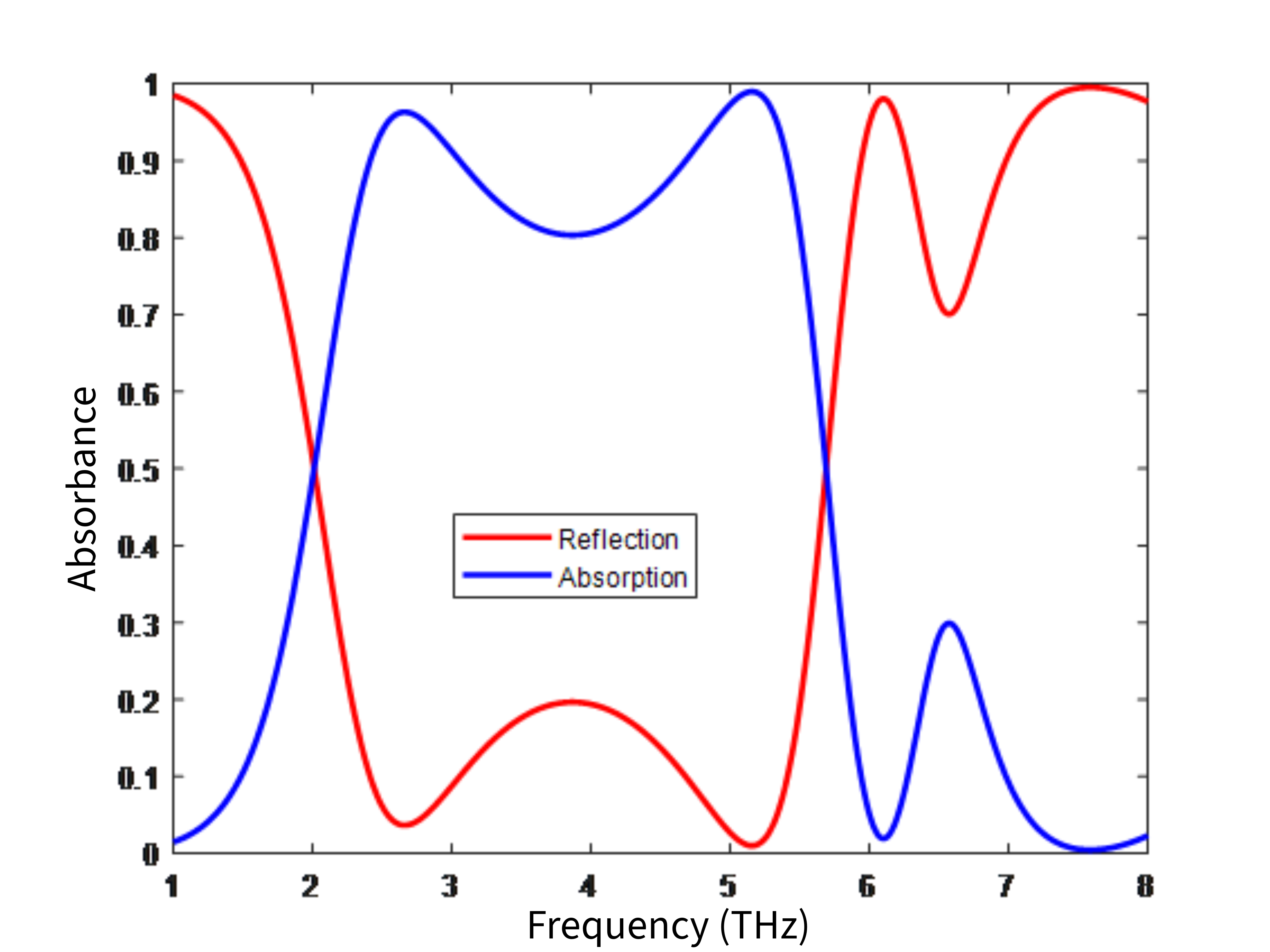}}
\caption{Absorption and reflection spectrum of the proposed model.}
\label{fig}
\end{figure}

\begin{figure}[htbp]
\centerline{\includegraphics[width=.9\linewidth]{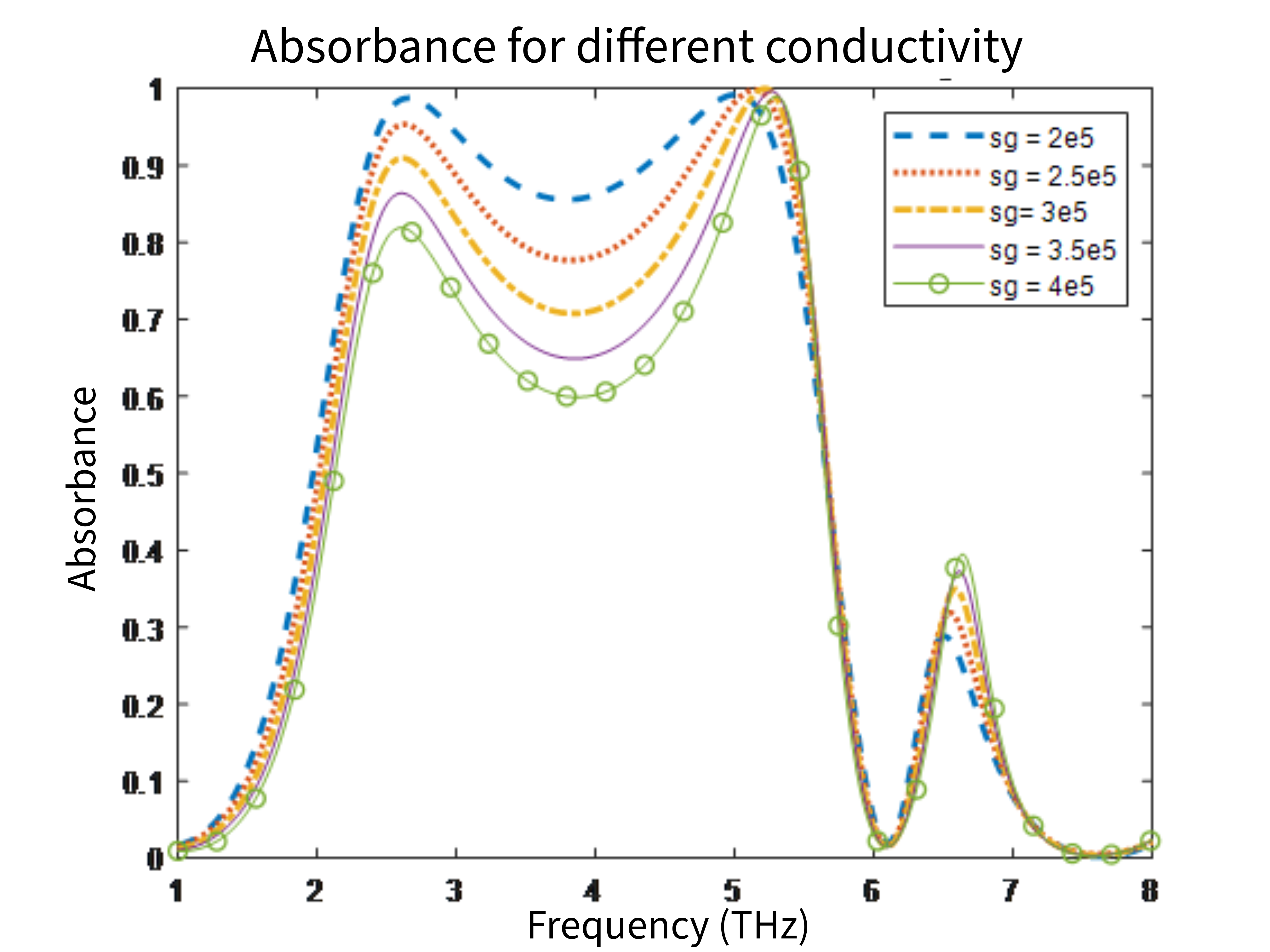}}
\caption{Absorbance for different conductivity of VO\textsubscript{2}.}
\label{fig:conductivity}
\end{figure}

\begin{figure}[htbp]
\centerline{\includegraphics[width=\linewidth]{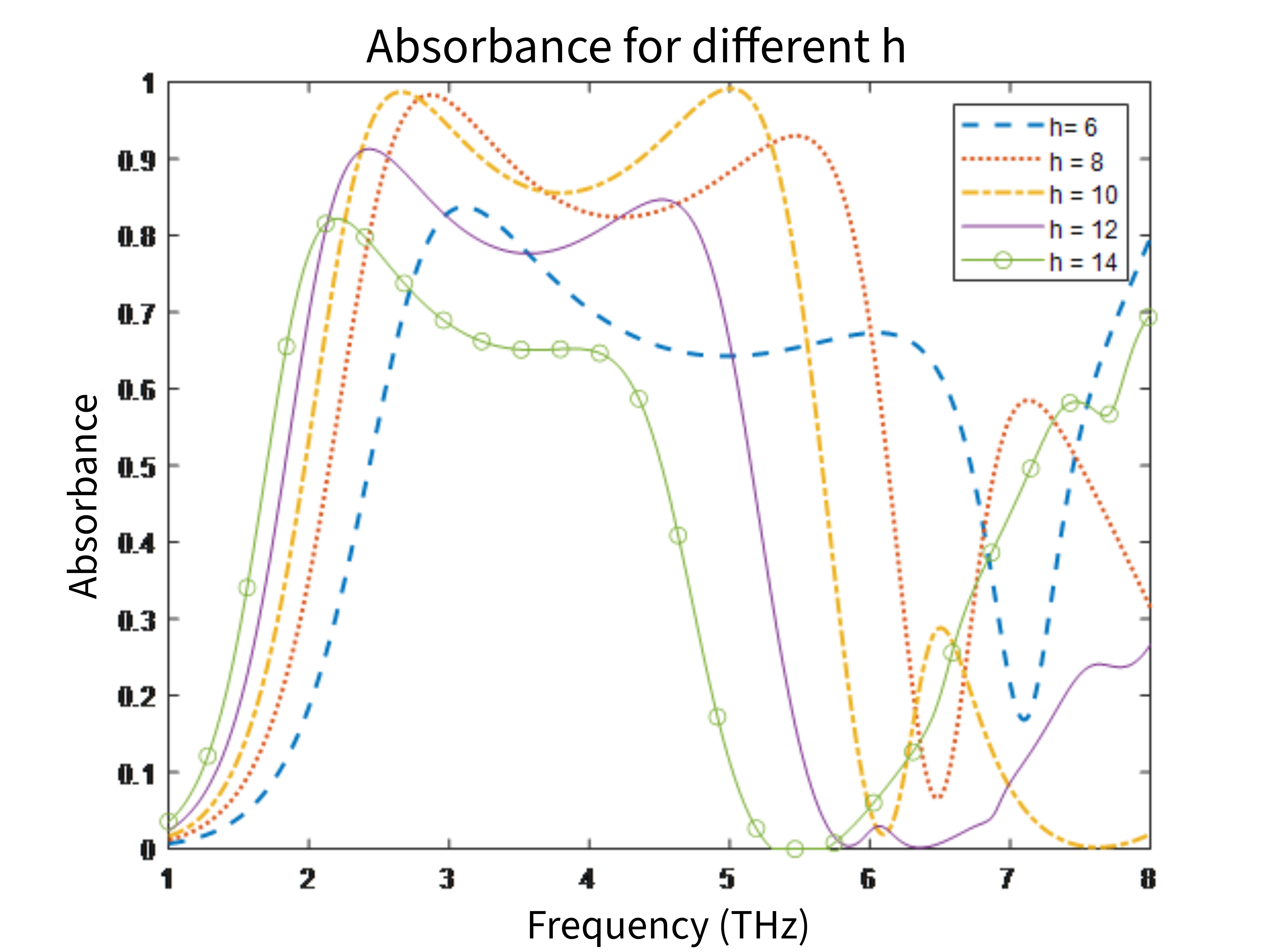}}
\caption{Absorbance for various SiO\textsubscript{2} thickness level.}
\label{fig:thickness}
\end{figure}

\begin{figure}[htbp]
\centerline{\includegraphics[width=\linewidth]{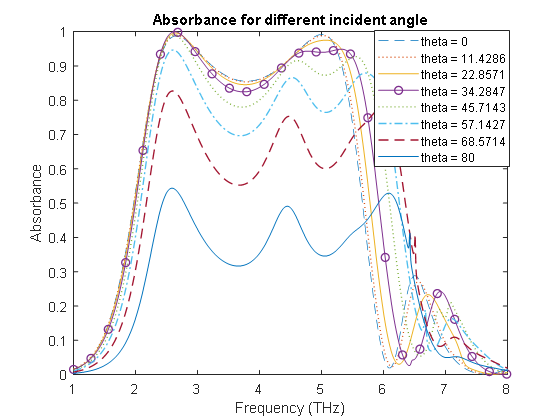}}
\caption{Absorbance for different incident angle (Theta).}
\label{fig:incident}
\end{figure}

\begin{figure}[htbp]
\centerline{\includegraphics[width=\linewidth]{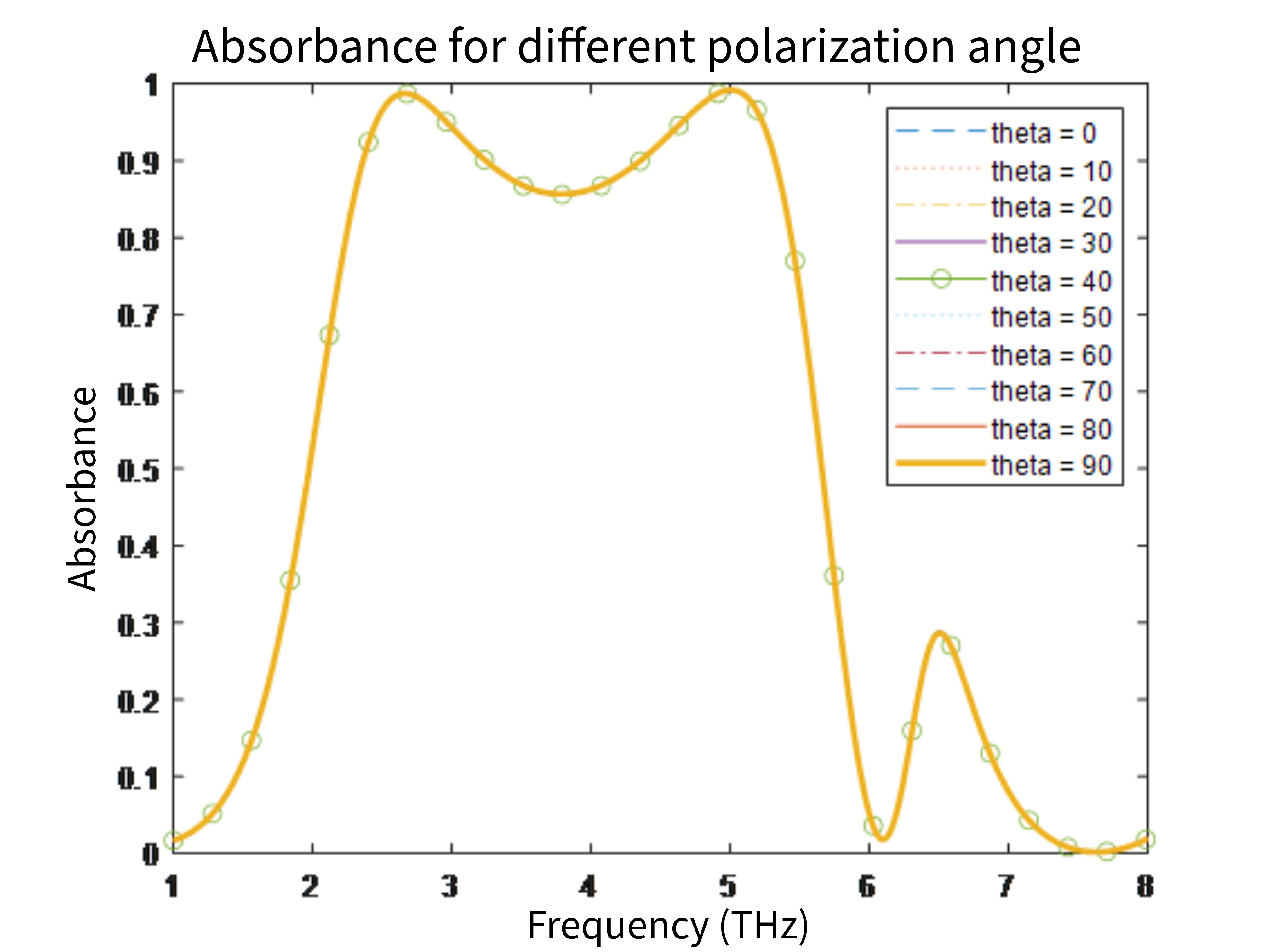}}
\caption{Absorbance for different polarization angles.}
\label{fig:polarization}
\end{figure}

The simulation outcomes indicate that the proposed model attains a wide absorption bandwidth of 3.00 terahertz (THz), from 2.414 THz to 5.417 THz.  Throughout this range, the model consistently achieves an absorption efficiency surpassing 90\%, as illustrated in Fig.~\ref{fig}.  Notably, the highest absorption efficiency is achieved at frequencies of 2.638 THz and 5.158 THz, where two distinct absorption peaks are observable. The central frequency of the model is positioned at 3.84 THz. This broad bandwidth and high absorption make the proposed absorber highly effective for applications across the terahertz spectrum.

As seen in Fig.~\ref{fig:conductivity}, the absorption efficiency decreases with increasing conductivity of VO\textsubscript{2}. Higher conductivity enables more efficient electron movement, which reduces the material’s ability to retain absorbed energy as heat. This tunability of the absorption with varying conductivity is a key feature, allowing for dynamic control of the metamaterial’s performance. The tunable nature of VO\textsubscript{2} provides flexibility in optimizing the absorber for different operational conditions.

Fig.~\ref{fig:thickness} presents the effect of varying the thickness of the SiO\textsubscript{2} layer (h) on the absorption performance. The thickness was varied from 6~\textmu m to 14~\textmu m. It is evident that the optimal thickness for achieving maximum absorption is $h = 10~\mu\text{m}$. Beyond this value, the absorption efficiency decreases, indicating that the thickness of the dielectric layer plays a crucial role in determining the overall performance of the metamaterial absorber.

In Fig.~\ref{fig:incident},  the absorption decreases with increasing incident angle. As the incident angle increases, the absorption in materials typically decreases due to changes in the path and penetration depth of light within the material. At steeper angles, light interacts less effectively with the absorbing elements of the material, as the effective thickness through which light travels increases, reducing the intensity of absorption. This phenomenon is often observed in thin films and coatings, where the orientation of the material relative to the light source critically affects the absorbance 
efficiency.

The graph in Fig.~\ref{fig:polarization} illustrates the absorption behavior of the model for different polarization angles. The results show that the absorption remains nearly constant, regardless of the polarization angle. This demonstrates the polarization-insensitive nature of the proposed absorber, which is crucial for practical applications where the orientation of the incident wave may vary. The symmetry of the metamaterial ensures consistent absorption performance independent of polarization.

\begin{table*}[h]
\centering
\caption{Comparison of the Proposed Model with Existing Designs}
\resizebox{\textwidth}{!}{
\begin{tabular}{|p{3cm}|c|c|p{5.5cm}|}
\hline
\textbf{References} & \textbf{Absorption Rate} & \textbf{Bandwidth (THz)} & \textbf{Material and Design Approach} \\
\hline
Wang \textit{et al.}, 2019 \cite{wang2019vanadium}  & >90\%    & 0.65 & VO\textsubscript{2}-based, Moderate Bandwidth \\
\hline
Bai \textit{et al.}, 2019 \cite{bai2019tunable}  & >95\%    & 1.25 & VO\textsubscript{2}-based, Multi-Layered \\
\hline
Chen \textit{et al.}, 2020 \cite{chen2020broadband} & >90\% & 2.7 & Graphene-based, Complementary Gammadion Shape, Single-Layer Tunable \\
\hline
Xie \textit{et al.}, 2021 \cite{xie2021tunable} & >95\% & 1.6  & Graphene-based, Fractal Metasurface, Broadband Absorber \\
\hline
Elkorany \textit{et al.}, 2023 \cite{elkorany2023design} & 100\% & 0.25 & Metallic, Polarization-Insensitive, Narrowband \\
\hline
Ri \textit{et al.}, 2024 \cite{ri2024tunable} & ~90\% & 2.18 & Graphene-based, Complementary Split-Ring, High Tunability \\
\hline
Zhang \textit{et al.}, 2024 \cite{zhang2024graphene} & 92\% & 2.5 & Metal-Dielectric, Frequency-Amplitude Tunable \\
\hline
\textbf{This Model} & ~99\% & 3.00 & VO\textsubscript{2}-based, Polarization-Insensitive, Broadband,  High-Efficiency \\
\hline
\end{tabular}}
\label{table:comparison}
\end{table*}

\begin{figure}[htbp]
\centerline{\includegraphics[width=.8\linewidth]{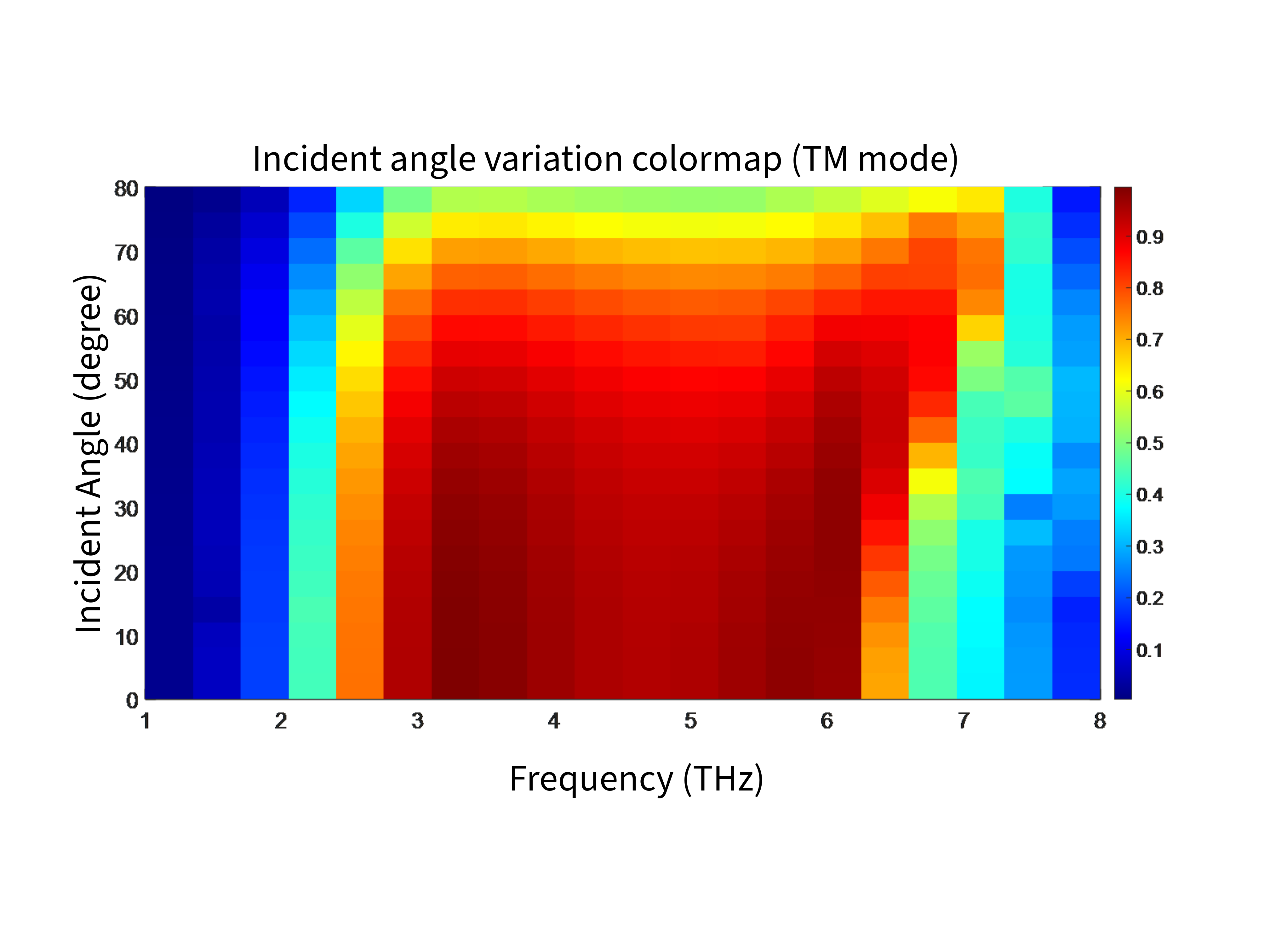}}
\caption{Effect of varying incident angles on the absorption.}
\label{fig:incident_angle}
\end{figure}

\begin{figure}[htbp]
    \centering
    \begin{minipage}{0.32\columnwidth}
        \centering
        \includegraphics[width=\linewidth]{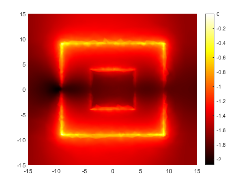}
        \caption*{(a)}  
    \end{minipage}
    \hfill
    \begin{minipage}{0.32\columnwidth}
        \centering
        \includegraphics[width=\linewidth]{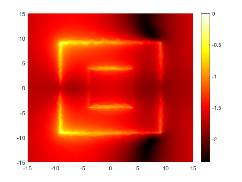}
        \caption*{(b)}  
    \end{minipage}
    \hfill
    \begin{minipage}{0.32\columnwidth}
        \centering
        \includegraphics[width=\linewidth]{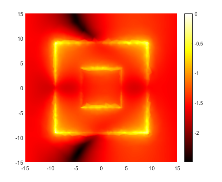}
        \caption*{(c)}  
    \end{minipage}
    \caption{Electric field distribution at (a) 2.64 THz, (b) 3.84 THz, and (c) 5.15 THz.}
    \label{fig:three_images}
\end{figure}

Fig.~\ref{fig:incident_angle} displays a colormap showing the effect of varying incident angles on the absorption efficiency. The results indicate a significant change in the absorption spectrum with variations in the incident angle. The ability of the metamaterial to maintain high absorption across different angles of incidence makes it suitable for wide-angle applications, such as in sensing and imaging devices.

Fig.~\ref{fig:three_images} illustrates the electric field distribution at the peak absorption frequencies of 2.64 THz and 5.15 THz. And also at the centre frequency of 3.84 THz. The field distributions are highly concentrated within the metamaterial structure at peak frequencies, indicating that the structure and the incoming electromagnetic waves are experiencing intense resonant interactions.  The efficient trapping of electromagnetic energy is emphasized by the field concentrations, which leads to an increase in absorption.  The center frequency, 3.84 THz, is not associated with a peak absorption point; rather, it is situated between the two peak.  The field distribution at this frequency may be less intense or exhibit a distinct pattern in comparison to the peak frequencies.  This suggests that the absorber is less effective in capturing energy at this frequency.  In comparison to the center frequency, the intensity of the electromagnetic fields within the metamaterial is stronger at peak frequencies (2.64 THz and 5.15 THz).  This suggests a more effective energy absorption and a stronger resonance.   These patterns also indicate the existence of electric dipole and magnetic dipole resonances, which are responsible for the metamaterial's exceptional absorption capabilities. s.

\subsection{Comparative Study}

Table~\ref{table:comparison} presents a detailed comparative analysis of the proposed metamaterial absorber in relation to existing designs. The findings reveal that the proposed model outperforms prior designs in terms of absorption efficiency, bandwidth, and structural adaptability.

Metallic absorbers, such as \cite{elkorany2023design}, achieve perfect absorption but are restricted to a narrow operational bandwidth of 0.25 THz. Their polarization insensitivity is a strength, but the limited spectrum restricts their broader applicability.

VO\textsubscript{2}-based absorbers \cite{wang2019vanadium}, and \cite{bai2019tunable} offer moderate bandwidths (0.65 and 1.25 THz), relying on phase-transition properties for tunability. However, they either exhibit complex multi-layered structures \cite{bai2019tunable}.

Graphene-based absorbers, including \cite{ri2024tunable}, \cite{zhang2024graphene}, \cite{xie2021tunable}, and \cite{chen2020broadband}, introduce electrical tunability. \cite{xie2021tunable} demonstrates fractal metasurface-based broadband absorption with high efficiency (>95\%), but its bandwidth (1.6 THz) remains lower than that of the proposed model. \cite{chen2020broadband} utilizes a complementary gammadion-shaped graphene design achieving 2.7 THz bandwidth, making it a strong candidate for tunable THz applications.

While these graphene-based models offer promising tunability, they still fall short of achieving the broadband, high-efficiency performance observed in the proposed absorber. The proposed absorber distinguishes itself by combining high efficiency (~99\%) with a broadband response (3.00 THz). Unlike graphene-based designs that require geometric alterations for tunability, this model leverages VO\textsubscript{2}'s phase transition properties for dynamic absorption control without modifying structural parameters. The simplicity of its fabrication, combined with its superior broadband performance, positions it as a highly efficient candidate for solar energy harvesting.

\section{Conclusion}

This paper has presented the development of a high-performance metamaterial absorber designed to achieve exceptional absorption efficiency in the terahertz (THz) spectrum. The proposed absorber demonstrates a remarkably broad absorption bandwidth of 3.00 THz, ranging from 2.414 THz to 5.417 THz, with a near-unity peak absorption of approximately 99\%.

The results of this study highlight the absorber's potential for a wide array of applications, particularly in THz imaging, sensing, and energy harvesting. Its tunability, achieved through the phase transition properties of vanadium dioxide (VO\textsubscript{2}), enables dynamic control over absorption characteristics, making it highly adaptable for reconfigurable devices. Additionally, the polarization-insensitive nature of the design ensures robust performance regardless of incident wave orientation, further enhancing its practical applicability.

As THz technology continues to advance, the proposed metamaterial absorber offers a promising solution for next-generation photonic and optoelectronic applications. Its combination of broadband response, high efficiency, and structural simplicity positions it as a viable candidate for integration into future THz-based communication systems, security screening technologies, and biomedical diagnostics. Future research could further optimize the absorber’s tunability and explore its implementation in real-world experimental setups, paving the way for practical deployment in diverse THz applications.

\bibliographystyle{IEEEtran}
\bibliography{References}

\end{document}